\documentclass[%
 reprint,
 amsmath,amssymb,
 aps,
]{revtex4-2}

\usepackage{graphicx}
\usepackage{dcolumn}
\usepackage{bm}
\usepackage{hyperref}

\begin{document}

\preprint{APS/123-QED}

\title{Factors influencing the energy gap in topological states of antiferromagnetic MnBi$_2$Te$_4$}

\author{A.~M.~Shikin}
\email{ashikin@inbox.ru}
 \affiliation{Saint Petersburg State University, 198504, St. Petersburg, Russia}
 
\author{T.~P.~Makarova}

\affiliation{Saint Petersburg State University, 198504, St. Petersburg, Russia}

\author{A. V. Eryzhenkov}
\affiliation{Saint Petersburg State University, 198504, St. Petersburg, Russia}

\author{D.~Yu.~Usachov}
\affiliation{Saint Petersburg State University, 198504, St. Petersburg, Russia}

\author{D.~A.~Estyunin}
\affiliation{Saint Petersburg State University, 198504, St. Petersburg, Russia}

\author{D.~A.~Glazkova}
\affiliation{Saint Petersburg State University, 198504, St. Petersburg, Russia}

\author{I.~I.~Klimovskikh}
\affiliation{Saint Petersburg State University, 198504, St. Petersburg, Russia}

\author{A. G. Rybkin}
\affiliation{Saint Petersburg State University, 198504, St. Petersburg, Russia}

\author{A.~V.~Tarasov}
\affiliation{Saint Petersburg State University, 198504, St. Petersburg, Russia}

\begin{abstract}
The experimentally measured angle-resolved photoemission dispersion maps for MnBi$_{2}$Te$_{4}$ samples, which show different energy gaps at the Dirac point (DP), are compared with the results of theoretical calculations to find the conditions for the best agreement between theory and experiment. We have analyzed different factors which influence the Dirac gap width: (i) the surface van der Waals (SvdW) distance between the first and second septuple layers (SLs), (ii) the magnetic moment on Mn atoms, (iii) the spin-orbit coupling (SOC) strength for the surface Te and Bi atoms and related changes in the localization of the topological surface states (TSSs). It was shown that all these factors may change the gap width at the DP in a wide range from 5 to $\sim$90~meV. We show that the Dirac gap variation is mainly determined by the corresponding changes in the TSSs spatial distribution. The best agreement between the presented experimental data (with the Dirac gaps between $\sim$15 and 55~meV) and the calculations takes place for a slightly compressed SvdW interval (of about -3.5~\% compared to the bulk value) with modified SOC for surface atoms (that can occur in the presence of various defects in the near-surface region). We show that upon changing the values of the SvdW interval and surface SOC strength the TSSs spatial distribution shifts between the SLs with opposite magnetizations, which leads to a non-monotonic change in the Dirac gap size.   
\end{abstract}

\maketitle


\section{\label{sec:level1}Introduction}

Detailed analysis of the electronic and spin structure, as well as the magnetic and electronic transport properties of magnetic topological insulators (TIs) have recently attracted enhanced interest due to the unique combination of topology and magnetism, leading to a quantized magnetoelectric (ME) response characteristic of these materials \cite{PhysRevB.78.195424,RevModPhys.83.1057,Chang_2013,2019NatRP...1..126T,3fe956269f1c4d61b90f3b51ef63c5f2,PhysRevB.92.081107,doi:10.1126/science.aax8156,article,PhysRevB.101.205130,Coh_2011,Xu_2012,doi:10.1126/science.1189924,Checkelsky2012DiracfermionmediatedFI}. The most impressive manifestations of such effects are the quantum anomalous Hall effect (QAHE) and the topological quantum ME effect based on quantization of the ME response and Hall conductivity (see, for example, articles \cite{PhysRevB.78.195424,RevModPhys.83.1057,Chang_2013,2019NatRP...1..126T,3fe956269f1c4d61b90f3b51ef63c5f2}). The formed non-trivial topological properties make magnetic TIs extremely attractive for applications in modern nanoelectronics and emerging 2D \cite{Gibertini_2019,article1,article2} and antiferromagnetic spintronics \cite{Jungwirth_2016,2018NatPh..14..242S,RevModPhys.90.015005}.

Intensive study of these effects began with the discovery of TIs doped with magnetic impurities \cite{PhysRevB.78.195424,RevModPhys.83.1057,Chang_2013,2019NatRP...1..126T,3fe956269f1c4d61b90f3b51ef63c5f2,PhysRevB.92.081107,doi:10.1126/science.aax8156,article,PhysRevB.101.205130,Coh_2011,Xu_2012,doi:10.1126/science.1189924,Checkelsky2012DiracfermionmediatedFI}. Magnetic impurities in TIs violate the time reversal symmetry, which leads to formation of a nonzero effective mass for the TSSs quasiparticles and the opening of an energy gap at the DP \cite{Xu_2012,doi:10.1126/science.1189924,Checkelsky2012DiracfermionmediatedFI, PhysRevB.97.245407,article_shikin,PhysRevLett.108.206801, article4} with its size proportional to the developed magnetic moment acting on the TSSs. Last time, significant efforts have been directed to the study of intrinsic magnetically ordered TIs, where magnetic atoms are directly included in the chemical structure of the synthesized magnetic TI. This provides an ordered arrangement of magnetic atoms inside the crystal lattice and allows a significant increase in the concentration of magnetic atoms and corresponding size of the gap open at the DP. One of such intensively studied intrinsic magnetic TIs is the antiferromagnetic  (AFM) TI with stoichiometry MnBi$_{2}$Te$_{4}$ (see, for instance, \cite{Otrokov24,zhang2019topological,li2019intrinsic,gong2019experimental,lee2019spin,aliev2019novel}), which, according to theoretical estimates, is characterized by the energy gap at the DP up to 88~meV \cite{Otrokov24}. Additionally, for this material, a possibility of realizing the quantum Hall effect was shown, both theoretically \cite{otrokov2019unique,liu2020robust} and experimentally \cite{doi:10.1126/science.aax8156,liu2020robust,ge2020high}, and for thin MnBi$_{2}$Te$_{4}$ layers the QAHE \cite{doi:10.1126/science.aax8156} was recently realized. These facts greatly increase the applied interest to the study of the electronic and magnetic properties of this material. 

The AFM TI MnBi$_{2}$Te$_{4}$ is a layered compound based on the Bi$_{2}$Te$_{3}$ quintuple layer~(QL) with an ordered MnTe magnetic bilayer embedded inside \cite{Otrokov24,zhang2019topological}, which leads to formation of the septuple layer~(SL) block with a well-developed ferromagnetic (FM) interaction between Mn atoms inside the magnetic Mn layer. The interaction between neighboring Mn layers has an AFM character \cite{Otrokov24,zhang2019topological}. As a result, AFM TI MnBi$_{2}$Te$_{4}$ is a layered compound consisting of a sequence of the Te-Bi-Te-Mn-Te-Bi-Te SL blocks separated by van der Waals (vdW) intervals \cite{aliev2019novel,yan2019crystal,zeugner2019chemical,li2020competing}. According to the theoretical estimates, the energy gap width at the DP for MnBi$_{2}$Te$_{4}$ is about 80–88~meV \cite{Otrokov24,zhang2019topological,shikin2022modulation}. However, the experimental studies show that the gap size varies in a wide range mostly from 50–70~meV to 12–15~meV and even smaller \cite{Otrokov24,lee2019spin,estyunin2020signatures,shikin2020nature,shikin2021sample,shikin2022modulation}.  Moreover, a number of works have appeared in the literature \cite{hao2019gapless,chen2019topological,swatek2020gapless,nevola2020coexistence,yan2021origins}, in which the possibility of a “gapless” TSSs dispersion was shown by angle-resolved photoemission spectroscopy (ARPES). At the same time, the recent works \cite{shikin2020nature,shikin2021sample,shikin2022modulation} show that the energy gap at the DP can vary significantly for different samples depending on the photon energy and other factors, such as defectiveness of samples, local variation in stoichiometry, modification of surface magnetic ordering, and etc.

For explanation of these differences a number of ideas have been put forward. In particular, in a number of works the "absence" of the energy gap at the DP was associated with the modification of the magnetic order in the upper surface layer \cite{hao2019gapless,bernevig2022progress}. In this case, according to theoretical modeling \cite{hao2019gapless}, it is assumed that the formation of the gapless dispersion can occur due to magnetic reconstruction of the surface, which leads to an effective decrease in the out-of-plane component of the magnetic moment in the surface Mn-layer. This may be due to the following reasons: (a) - formation of a local AFM coupling along the surface in the surface magnetic layer, (b) - rotation of the magnetic moments in the surface magnetic layer along the surface, and (c) - formation of a magnetically disordered paramagnetic or a “dead” magnetic layer on the surface, which leads to the closing of the Dirac gap \cite{hao2019gapless}. At the same time, according to theoretical estimates \cite{shikin2020nature}, the size of the Dirac gap for the MnBi$_{2}$Te$_{4}$ can also be changed due to the surface relaxation of the vdW distances between the first and second upper SLs \cite{shikin2020nature}). The resulting shift of the TSSs localization towards the second lower-lying SL, which is characterized by the opposite orientation of the Mn magnetic moments, can lead to compensation of the effective out-of-plane magnetic moment in the TSSs localization region and the corresponding shrinking of the Dirac gap. Similarly, a change in the Dirac gap size can occur due to the presence and accumulation of various surface defects of different concentrations \cite{shikin2021sample,shikin2022modulation,garnica2022native}, including possible adsorption of residual gas molecules, which can cause the changes in the TSSs localization resulting in modulation of the energy gap at the DP \cite{shikin2021sample,shikin2022modulation}.

It follows from the foregoing that the questions about the reasons for the modulation of the Dirac gap in MnBi$_{2}$Te$_{4}$ and its relation to the magnetic interactions remains open and require further analysis. In this work, we will try to analyze different factors responsible for changes in the electronic structure of the TSSs and the nearest conduction and valence bands (CB and VB) states in MnBi$_{2}$Te$_{4}$, as well as the corresponding modulation of the Dirac gap by varying the parameters of the theoretical model related to the magnetic and spin-orbit interactions. We will consider: (i) the role of variation in the SvdW interval between the first and second SLs, (ii) the effect of modulation of the magnetic moment on Mn atoms, (iii) the influence of the SOC strength for surface Te and Bi atoms, as well as the related redistribution in the localization of TSSs. The calculation results will be analyzed in comparison with the experimental dispersions measured by ARPES. The experimental basis for such variation of calculation parameters can be: (i) surface relaxation processes leading to a change in the SvdW interval \cite{shikin2020nature}, (ii) the formation of Mn/Bi type substitution defects, which reduce the exchange field in the region of localization of the TSSs \cite{garnica2022native}, and (iii) the formation of various types of defects in the near-surface region that may change the surface potential gradient, which can be taken into account by artificial variation of the SOC strength for the surface atoms \cite{shikin2021sample,shikin2022modulation}. At the same time, such a change in the parameters of the theoretical model can also include the situation of changing the composition of the sample surface by replacing some of the atoms in the near-surface layer with atoms possessing a lower atomic SOC, which may be important in the formation of synthetic layered topological structures.

\section{\label{sec:level2}Methods}
\subsection{ARPES measurements}
The measurements of the ARPES dispersion maps were carried at the $\mu$-Laser ARPES system at HiSOR (Hiroshima, Japan) with improved angle and energy resolution and a high space resolution of the laser beam (spot diameter around 5 $\mu$m) using a Scienta R4000 analyzer with an incidence angle of the LR of 50$^{\circ}$ relative to the surface normal at photon energy of 6.3~eV.
High-quality MnBi$_{2}$Te$_{4}$ single crystals were grown using the vertical Bridgman method at the Novosibirsk State University. Clean surfaces of the samples were obtained by a cleavage in ultrahigh vacuum. The base pressure during all photoemission experiments was better that $1\times 10^{-10}$~mbar.

\subsection{DFT calculations}
The electronic structure calculations were done by using OpenMX code, providing the fully relativistic DFT (density functional theory) implementation with localized pseudoatomic orbitals \cite{ozaki2003variationally,ozaki2004numerical,ozaki2005efficient} and the norm-conserving pseudopotential \cite{troullier1991efficient}. The exchange-correlation energy in PBE version of generalized gradient approximation was exploited \cite{perdew1996generalized}. The accuracy of the real-space numerical integration was specified by the cutoff energy of 450 Ry, the total-energy convergence criterion was $1\times 10^{-6}$~eV, whereas the surface Brillouin zone of the supercell was sampled with a $5\times 5$ mesh of k points.

The basic functions were taken as follows: Bi8.0 — s3p2d2f1, Te7.0 — s3p2d2f1, Mn6.0 — s3p2d1, this means the potential cutoff radius and a set of basic functions. The Mn~3d states were treated within the DFT + U approach \cite{han2006n} within the Dudarev scheme \cite{dudarev1998electron} where U parameter equals 5.4~eV \cite{Otrokov24}.
The surface was represented by a repetitive slab of 6~SLs of MnBi$_{2}$Te$_{4}$. The vacuum layer of 12~$\buildrel _{\circ} \over {\mathrm{A}}$ was placed between slabs to avoid their interaction.

\section{\label{sec:level2}Results and Discussion }
\subsection{Electronic structure of the TSS\lowercase{s} and the nearest VB and CB states for M\lowercase{n}B\lowercase{i}$_{2}$T\lowercase{e}$_{4}$ measured by ARPES}

\begin{figure*}
\includegraphics[width=0.65\linewidth]{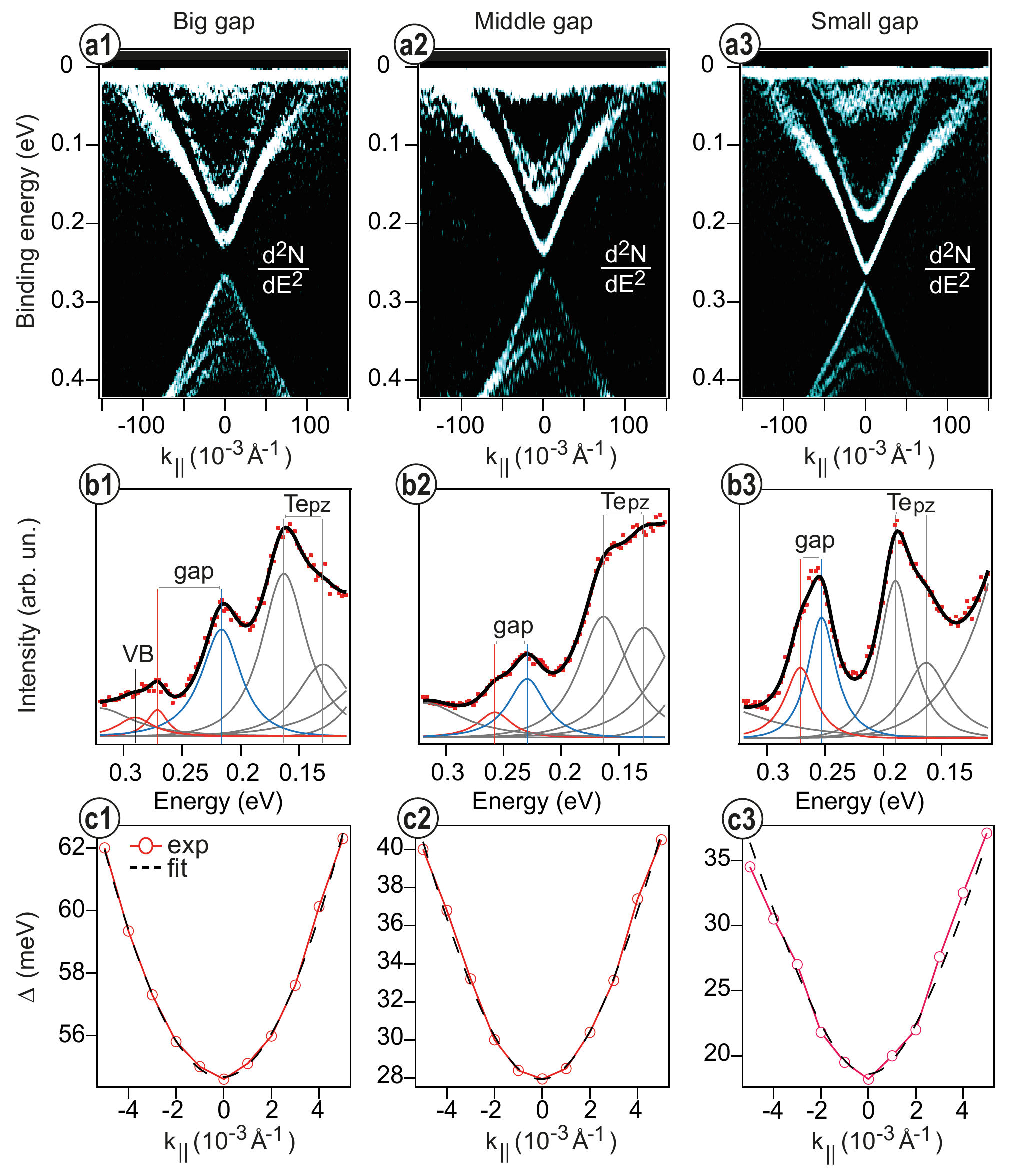}
\caption{\label{fig:fig1}(a1-a3) - Dispersion maps for the TSSs and nearest VB and CB states measured by ARPES for different samples of MnBi$_{2}$Te$_{4}$, which exhibit different gaps at the DP. The dependencies are presented in the form of second energy derivative for better visualization of the features in the electronic structure. (b1-b3) - Corresponding EDCs at $k_{||}=0$ with deconvolution into spectral components. (c1-c3) – Changes in the energy separation between the states of the upper and lower parts of the Dirac cone as a function of $k_{||}$. The smallest splitting corresponds to the Dirac gap size.}
\end{figure*}

Figures \ref{fig:fig1}(a1) -- (a3) show the ARPES dispersion maps showing the features of electronic structure of the TSSs for several samples of MnBi$_{2}$Te$_{4}$ differing in the experimentally measured gap sizes at the DP. The spectra were measured at a temperature of 10–15~K using laser radiation (LR) with photon energy of 6.3~eV. The inset (a1) shows the ARPES dispersions for the sample with a large gap of 54~meV at the DP. The insets (a2) and (a3) show similar ARPES maps taken from the samples with medium (28~meV) and small (18~meV) DP gaps. Such different experimentally measured Dirac gaps were also noted in Refs. \cite{shikin2020nature,shikin2021sample,shikin2022modulation}. In addition to the TSSs, the ARPES dispersion maps show the nearest VB and CB states. First of all, these are the exchange-split states with mainly Te~$p_z$ character, which are located in the CB at binding energies (BE) of about 0.13-0.18~eV. Their energy splitting is related to the magnetization of Mn layers, which depends on temperature \cite{estyunin2020signatures}. Experimental estimates of the Te~$p_z$ splitting for samples with a large gap, carried out in Ref.~\cite{estyunin2020signatures}, give values of 40–45~meV when extrapolated to zero temperature. Insets (b1-b3) show the corresponding spectra or energy distribution curves (EDCs) measured at the $\bar{\Gamma}$-point close to the DP. The lines of different colors represent the deconvolution into spectral components in the regions of the TSSs and the Te~$p_z$ states (blue/red and black peaks, respectively). The energy positions of the spectral components corresponding to the TSSs make it possible to estimate the Dirac gap size and its changes for different samples (see the vertical lines at the peak maxima).

Insets (c1-c3) show the corresponding changes in the energy separation ($\Delta$) between the states of the upper and lower parts of the Dirac cone (obtained from deconvolution of spectra) as a function of $k_{||}$. The presented dependencies were approximated by the equation: $\Delta(k_\|) \sim \sqrt{\alpha k_{||}^2 + \Delta_0^2}$, where $\Delta_0$ is the DP gap size, which was derived from the model dispersion for the massive Dirac quasiparticles. This approach allowed us to reduce the error in estimating the size of the gap. The approximations shown in Fig. \ref{fig:fig1}(c1-c3) resulted in the Dirac gap sizes for the presented cases of the large, middle and small gaps on the level of 54, 28 and 18~meV, respectively. The energy splitting of the Te~$p_z$ states seems to be comparable for all three samples. Although for the sample with a small Dirac gap one can distinguish a possible slight decrease in the splitting value.

\subsection{Theoretical calculations}
Below, we present the results of calculations of changes in the electronic structure of the TSSs and the nearest VB and CB states with variations in (i) the SvdW interval, (ii) the effective magnetic moment in the first two surface SLs (where the TSSs are mainly located), and (iii) the SOC strength for Te and Bi atoms in the region of the first two SLs.

The calculations in the present work were based on the volume unit cell used in Ref. \cite{Otrokov24}  ($a$ = 4.3336~\AA, $c$ = 40.959~\AA). We kept the atomic positions unchanged, however, optimization of the lattice parameters was performed to minimize the energy. The optimized parameters are characterized with a slightly increased (by 1~\%) value of $c$ and minor change of $a$ ($a$ = 4.3148~\AA, $c$ = 41.5444~\AA). The calculation details are given in the Methods section. The unit cell for the surface calculation is shown in Fig. 3S of Suppl. Mater. Our optimized cell results in a more pronounced cone-like shape of the lower part of the Dirac cone states in comparison with the results of calculations in Refs.~\cite{Otrokov24,shikin2020nature,shikin2021sample,shikin2022modulation} and smaller width of the gap open at the DP. As shown in Fig.~2S Suppl. Mater, gap value of 85~meV, similar to the value in Refs. \cite{Otrokov24,shikin2020nature,shikin2021sample,shikin2022modulation}, can indeed be obtained if we take the cell parameters from Ref.~\cite{Otrokov24}. However, as we already noted, at these parameters, the dispersion of the lower part of the Dirac cone states near the $\bar{\Gamma}$-point takes on a more plateau-like form.

\subsubsection{Influence of the SvdW interval on the Dirac gap width}

\begin{figure*}
\includegraphics[scale=0.7]{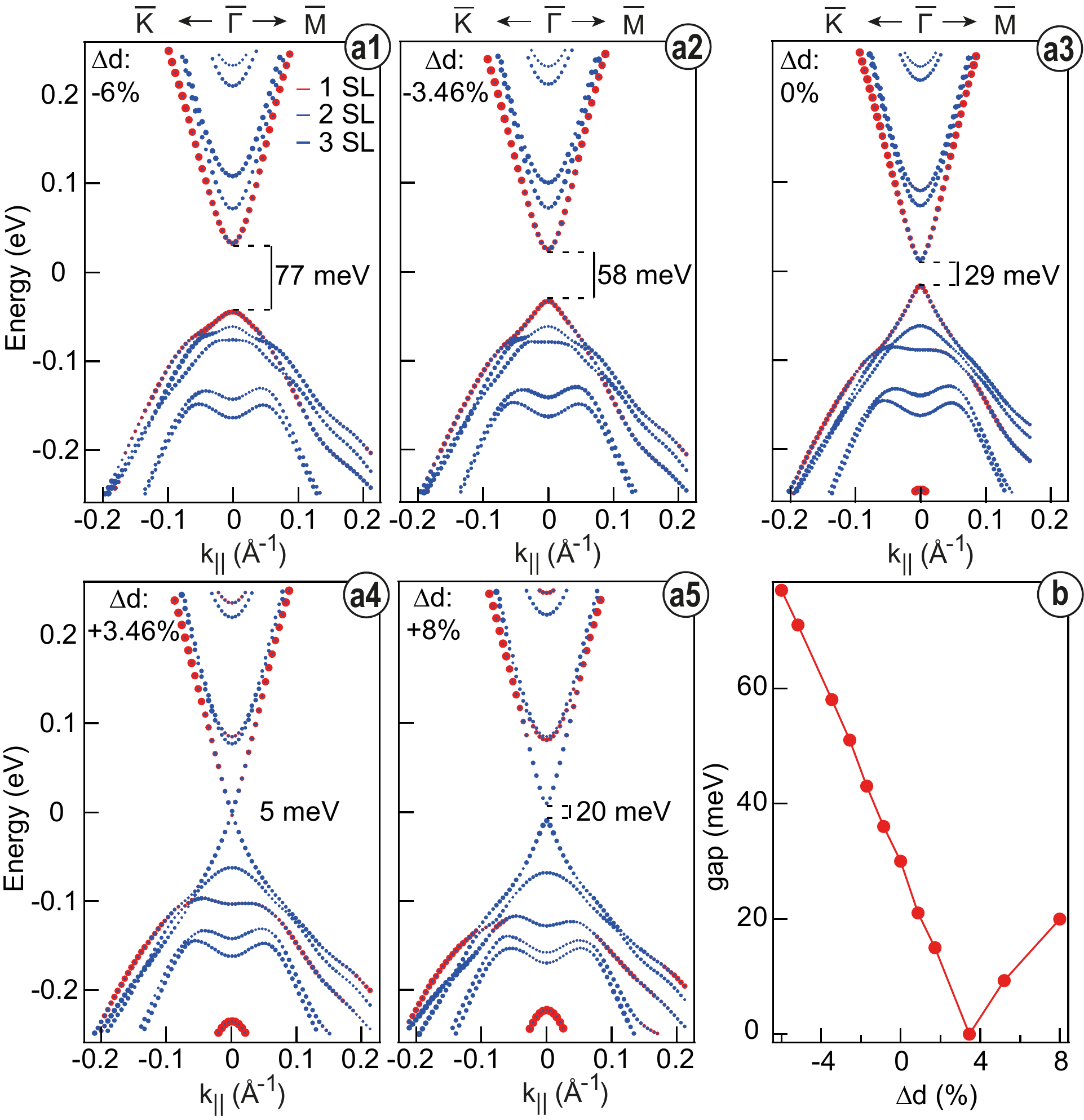}
\caption{\label{fig:fig2}(a1-a5) - Calculated electronic structure of the TSSs and the nearest VB and CB states for different vdW spacing between the first and second SLs ($\Delta$d~(\%)) with the indication of states localized in the first and second/third SLs (red and blue symbols, respectively). (b) - Dependence of the Dirac gap size on the relative change of the vdW distance between the first and second SLs ($\Delta$d~(\%)).}
\end{figure*}

 In the near-surface region, the SvdW interval can be changed due to surface relaxation, \cite{Otrokov24,shikin2020nature,shikin2021sample}. To study the influence of such structural changes on the size of the Dirac gap and on the electronic structure of the TSSs and the nearest CB and VB states, we performed DFT calculations for different SvdW distances between the first and second SLs (changed by $\Delta$d\%). The results are presented in Figs.~\ref{fig:fig2}(a1-a5), where the changes in the electronic structure are shown both with a decrease (to -6~\%), and an increase (to +8~\%) in the SvdW interval relative to the distance in the bulk (2.896~\AA), which is marked as 0~\%. A more complete set of the calculated dispersions in a wider energy range of the SvdW interval variation is presented in Fig.~1S Suppl. Mater.  
 
If we compare the experimental bands in Fig.~\ref{fig:fig1} and in Refs.~\cite{zhang2019topological,li2019intrinsic,estyunin2020signatures,shikin2020nature,shikin2021sample} with our calculation results, we see that the electronic structure calculated for the system with the SvdW interval compressed by -3.46\%, which provides the size of the DP gap of 58 meV (Fig.~\ref{fig:fig2}(a2)), is most consistent with the experimental dispersion maps for the case of a large gap presented in Fig.~\ref{fig:fig1}(a). When the SvdW interval is compressed by (-6~\%) (under maintaining the bulk vdW interval), the calculated gap size increases to 77~meV (see Fig.~\ref{fig:fig2}(a1)) that is close to that calculated in \cite{Otrokov24,shikin2020nature,shikin2021sample,shikin2022modulation}. As the SvdW interval expands, the size of the Dirac gap, on the contrary, decreases down to the minimal value of about 5 meV at the SvdW interval increased by +3.46~\% (see Fig.~\ref{fig:fig2}(a4)), and then again increases to 20~meV with further expansion of the SvdW interval to +8~\% (see Fig.~\ref{fig:fig2}(a5)). The complete calculated dependence of the Dirac gap width on the subsurface SvdW interval is shown in Fig.~\ref{fig:fig2}(b). A similar behavior of the Dirac gap size upon variation of the SvdW interval was presented in Ref.~\cite{shikin2020nature}, only with a slightly shifted minimum.

In addition to the TSSs dispersions Figs.~\ref{fig:fig2}(a1-a5) also show the the edge CB and VB states. One can see different exchange splittings of the Te~$p_z$ states in the CB that roughly correlate with the gap width. To separate the contributions of different SLs to the TSSs and the CB and VB states, the calculated states in Figs.~\ref{fig:fig2}(a1-a5) are marked by symbols of different colors, depending on their spatial localization. Red circles shows the states with the highest localization in the first surface SL, blue circles denote the states localized mainly in the second and third SLs. One can see that for the systems with large gap the TSSs are localized mainly in the region of the first SL (and partly in the deeper layers). Thus, from Figs.~\ref{fig:fig2}(a1-a5) it readily becomes apparent that with an increase in the SvdW interval the localization of the TSSs is shifting towards deeper SLs.


In contrast to the TSSs, the nearby CB and VB states do not show preferential localization in the first SL, pointing to their bulk-like nature. Figures~\ref{fig:fig2}(a1-a5) show that the increase of the SvdW distance is accompanied by a significant decrease in the energy splitting of the Te~$p_z$ states. Strong dependence of these states on the structural parameters near the surface indicates that they should be interpreted as surface resonances rather than purely bulk sates. Taking into account the surface sensitivity of ARPES, we suppose that these resonances should correspond to the Te~$p_z$ states observed in the experiment. In contrast to the slab calculations, where the splitting of the Te~$p_z$ states is notably different for the systems with different DP gap width, the experimental ARPES data do not show significant differences in the splitting of the Te~$p_z$ states among the samples with different DP gap. Based on the comparison of the calculated and experimentally observed splitting of Te~$p_z$ states and taking into account the fact that surface relaxation usually occurs in the inward direction, we exclude the cases of increased SvdW interval from further consideration. Thus, the experimental dispersions for the small (and probably middle) gap cannot be explained solely by the expansion of the SvdW interval. 

\begin{figure*}
\includegraphics[scale=0.7]{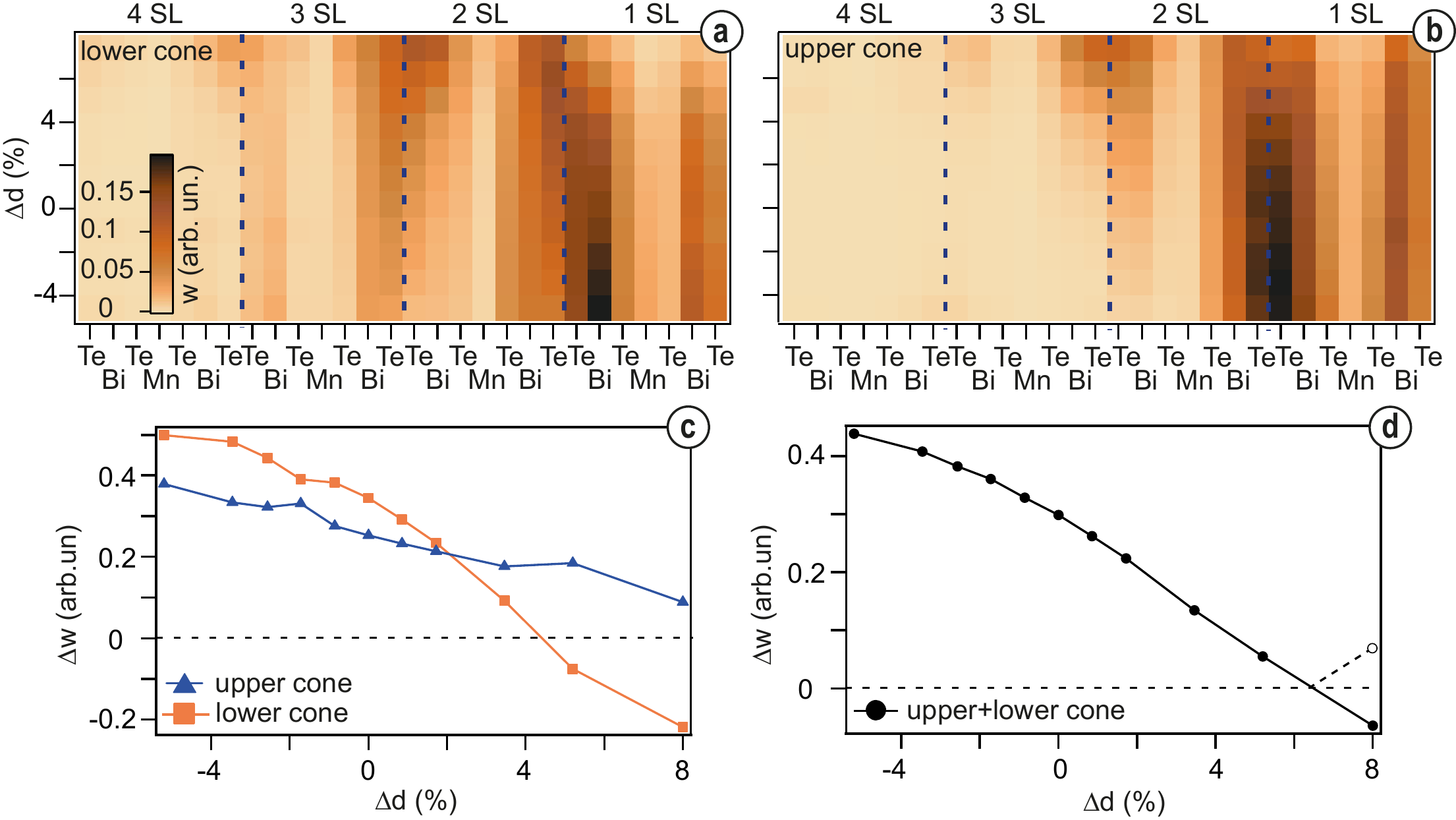}
\caption{(a,b) -- Redistribution in the TSSs local density for the lower and upper parts of the Dirac cone in the region of the upper four SLs upon variation of the vdW interval between the first and second SLs ($\Delta$d~(\%). High TSSs density  corresponds to dark color.  (c) -- Corresponding changes in the TSSs density multiplied by $-1^s$, where $s$ is the SL layer number, and averaged over four surface SLs for the states in the lower and upper parts of the Dirac cone (blue and orange symbols, respectively). (d) -- Corresponding changes in the TSSs density averaged over the upper and lower Dirac cone states.}
\label{fig:3}
\end{figure*}

Figures~\ref{fig:3}(a) and \ref{fig:3}(b) show the spatial distribution of the states of the lower and upper parts of the Dirac cone in the region of the four surface SLs as a function the SvdW distance between the first and second SLs. With an atomic orbital basis set the TSSs localization on every atom can be represented as the contribution of orbitals of a given atom $n$ to the TSSs at a given $k$-point: $w_k(n)=\sum\limits_{m}|\langle\phi_{nm}|TSS_k\rangle|^2$, where $|TSS_k\rangle$ is the TSSs and the summation runs over orbitals $\phi_{nm}$ of the considered atom $n$. The values of $w_k$ were averaged over three $k$-points closest to the $\bar{\Gamma}$-point in Fig.~\ref{fig:fig2}. The highest charge densities in the TSSs correspond to dark regions in Figs.~\ref{fig:3}(a,b). The presented TSSs distributions show that the TSSs are maximally localized at the edges of SLs. At the same time, the TSSs distribution is somewhat different for the states of the upper and lower parts of the Dirac cone. With a change in the SvdW interval, the localization of the TSSs for both parts of the Dirac cone changes, to a greater extent for the states of the lower part of the Dirac cone. The calculation shows that for the compressed SvdW interval (negative values of $\Delta$d), both the upper and lower parts of the Dirac cone turn out to be largely localized in the region of the first SL with the tails in the region of the second SL. With an increase of the SvdW distance, the TSSs change their localization towards the second and partially third SL. In this case, as it is shown in Fig.~\ref{fig:fig2}(b), the gap gradually decreases from 77~meV (at $\Delta$d = -6~\%) to 58~meV (at $\Delta$d = -3.46~\%) and down to 29~meV, when the SvdW interval becomes equal to the bulk value. With further expansion of the SvdW interval (by +3.46~\% with respect to the bulk), the TSSs both in the lower and upper parts of the Dirac cone become largely localized in the second SL and partially in the first and third ones. Once the TSSs become nearly equally localized in the first and second SLs, the gap width tends to the minimal value. With a further increase in the SvdW interval, the localization of the TSSs shifts more to the region of the third SL, and the gap width begins to increase again.

This analysis suggests that the changes in the size of the DP gap is mainly determined by the TSSs redistribution. When the TSSs are localized mainly in the first SL, the gap size is maximal. When the TSSs density shifts to region of the second SL, which is characterized by the opposite magnetic moments on atoms compared to the first SL, the exchange field acting on the TSSs decreases and the gap also reduces. With further shifting of the TSSs density towards the third SL, the size of the Dirac gap increases again. 

As some quantitative indicator of the influence of the TSSs localization on the DP gap width Fig.~\ref{fig:3}(c) shows the changes in the TSSs density averaged over four surface SLs (shown in Fig.~\ref{fig:3}(a)) for the states of the upper and lower parts of the Dirac cone (blue and orange symbols, respectively) upon variation of the SvdW interval between the first and second SLs. This averaged density was calculated based on the local density of states $w(n)$, taken with different signs for the SLs with opposite magnetizations. It is interesting that the presented dependencies show the change in the sign of the average TSSs density for the lower Dirac cone states, which are more sensitive to the changes of the SvdW interval. 
In this case, the zero-crossing point corresponds approximately to the minimum of the DP gap width in Fig.~\ref{fig:fig2}(b). Figure~\ref{fig:3}(d) demonstrates the changes in the total TSSs density averaged over the upper and lower Dirac cone contributions. This averaged dependence also shows a sign change, but with a slightly shifted zero-crossing point. 
The dotted line in Fig.~\ref{fig:3}(d) demonstrates changes in the absolute value of the signed TSSs density, which apparently correlates with the corresponding DP gap width. 

The presented dependencies confirm the fact that the DP gap modulation is indeed largely determined by the redistribution of the TSSs between different SLs characterized by oppositely directed magnetic moments. At the same time, the non-monotonic changes in the DP gap width can be related to the change in sign of the effective exchange magnetic field acting on the TSSs.

\subsubsection{\label{sec:citeref}Effect of magnetic moment on Mn atoms}

\begin{figure*}
\includegraphics[scale=0.6]{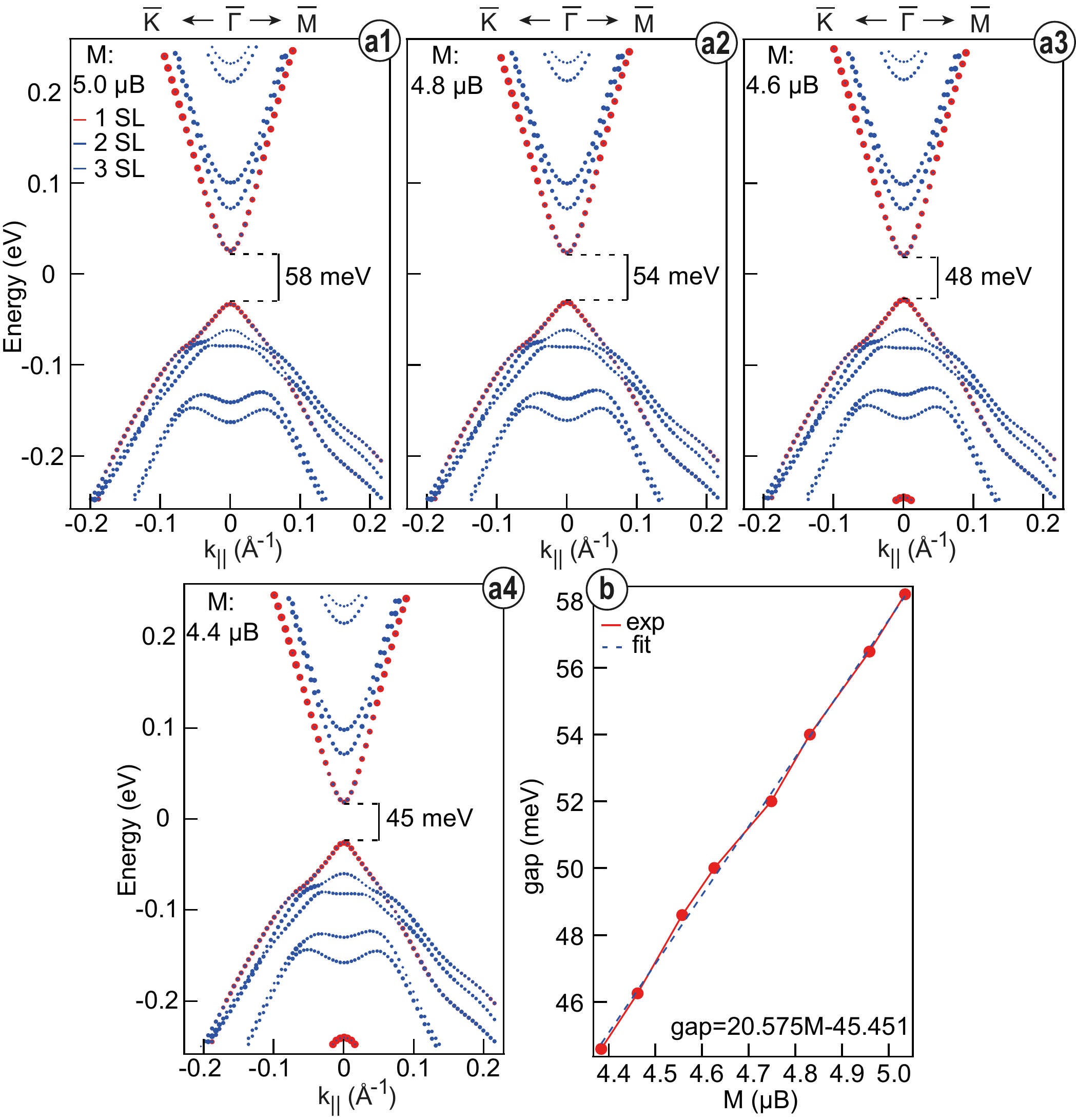}
\caption{\label{fig:fig4}(a) -- Calculated changes in the electronic structure of the TSSs and the nearest VB and CB states for MnBi$_{2}$Te$_{4}$ with indicated contributions of the states localized in the first (red symbols) and second/third (blue symbols) SLs upon variation of the magnetic moment on the Mn atoms. (b) -- Dependence of the energy gap at the DP on the magnetic moment on Mn atoms.}
\end{figure*}

In order to study the dependence of the DP gap width on the value of magnetic moments on Mn atoms, we carried out the calculations with constraint functional, which give a penalty unless the difference between the desired spin moment value and the initial one is zero \cite{kurz2004ab}. Figures~\ref{fig:fig4}(a1-a4) present the changes in the electronic structure of the TSSs, as well as the nearest VB and CB states in MnBi$_{2}$Te$_{4}$, with artificial variation (decrease) of the magnetic moments on Mn atoms relative to the value in the initial system. The system with the DP gap of 58~meV was taken as the initial one (see Fig.~\ref{fig:fig2}(a2)). In Figs.~\ref{fig:fig4}(a1-a4), as before, the contributions from the states localized in the first and second/third SLs are shown by red and blue symbols, respectively. 
It can be seen from the presented dispersion maps that as the magnetic moments on Mn atoms decrease from 5.035~$\mu_{B}$ to 4.38~$\mu_{B}$, the gap at the DP decreases from 58 to 45~meV. The dependence of the resulting gap size on changes in the magnetic moment on the Mn atoms is shown in Fig.~\ref{fig:fig4}(b). Interestingly, the continuation (approximation) of this dependence to a zero gap corresponds to a nonzero magnetic moment on the Mn atoms (approximately twice less than the initial value). No significant changes in the structure of the CB and VB states are observed. The electronic structure of the TSSs also remains almost unchanged. The top of the VB remains localized in the region of the lower cone of the Dirac states. Only the DP gap width changes.

The change in the effective magnetic moment can be developed, for instance, due to the substitution defects like Mn/Bi or Mn/Te \cite{garnica2022native}, when the magnetic Mn atoms are replaced by nonmagnetic ones. 
On the other hand, Mn atoms on the Te atomic sites have opposite magnetic moment orientation that also leads to a decrease of total exchange field acting on TSSs. 

\subsubsection{Effect of the spin-orbit coupling for Bi and Te surface atoms}

\begin{figure*}
\includegraphics[scale=0.5]{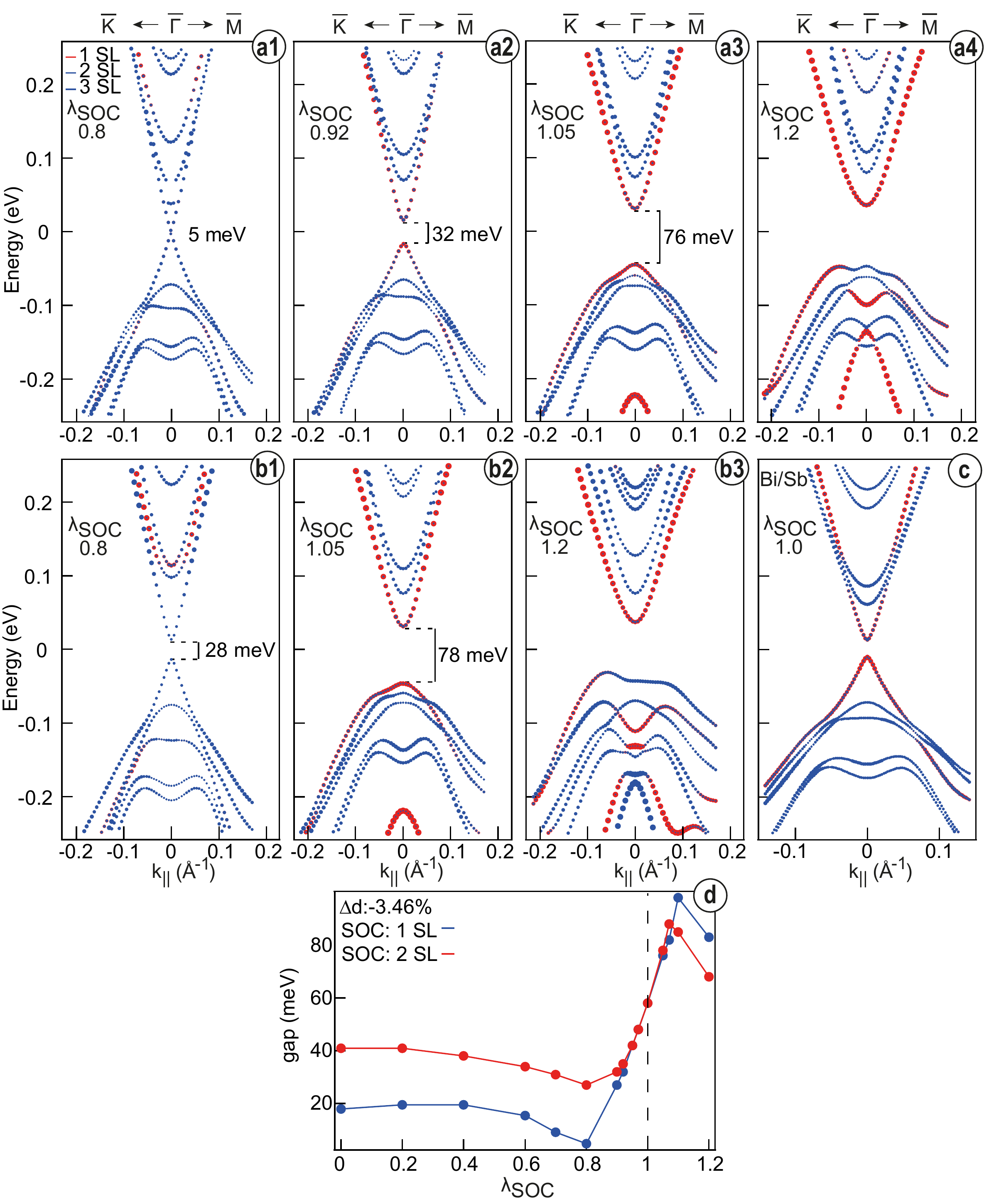}
\caption{\label{fig:5}(a1-a4, b1-b3) -- Calculated changes in electronic structure of the TSSs and the nearest VB and CB states for MnBi$_{2}$Te$_{4}$ under variation of SOC on the Te and Bi atoms in the first (a1-a4) and first two SLs (b1-b3) relative to the initial value taken as unity ($\lambda_{SOC}=1$). The system with the initial Dirac gap of 58~meV was used in calculations. The contributions of states localized in the first and second/third SLs are shown in red and blue symbols. (c) -- Changes in electronic structure under substitution of 50~\% of Bi-atoms by Sb- atoms in the first SL, for comparison.  (d) -- Corresponding change in the energy gap at the DP for both systems where SOC was varied in one or two SLs (SOC:1SL and SOC:2SL systems).}
\end{figure*}

In this part of the work, we analyze how the changes in the gap size at the DP and the corresponding changes in the electronic structure can be affected by the modulation of the effective SOC strength for surface atoms in the first two SLs. Such SOC modulation is aimed to describe the changes in the near-surface potential gradient and related TSSs localization due to the accumulation of various types of defects in the surface region (by analogy with Ref.~\cite{shikin2022modulation}) and possible substitution of the near-surface atoms by impurities with reduced atomic SOC. Wherein, we will describe and compare the effects of the SOC modulation both for atoms only in the first SL (in one case) and for atoms in two surface SLs (in the second case). This approach differs from the results in Ref. \cite{shikin2022modulation}, where the SOC strength was modulated only for the first two surface Te and Bi atomic layers. Here, we expanded the region of the surface SOC modulation to take into account the contributions of Te and Bi atoms near the SvdW interval. These layers presumably determine to a large extent the modulation of the DP gap  (see Fig.~\ref{fig:3} and Ref.~\cite{shikin2020nature}). Thus, the area of two surface SLs must be considered. In the current work the SOC near the surface was varied by scaling the SOC strength in all Te and Bi atoms of either one or two surface SLs. 

Figures~\ref{fig:5}(a1-a4) show the changes in the electronic structure of the TSSs, as well as the nearest VB and CB states under variation of the SOC strength parameter $\lambda_{SOC}$ in the range of 0.8--1.2 for Bi and Te atoms within the first surface SLs. The magnetic moments on the Mn atoms remained the same as for the initial system. The calculations were carried out for the system with the SvdW distance, changed by -3.46~\% (compressed) relative to the bulk value. The initial electronic structure (at $\lambda_{SOC}=1$), which is shown in Fig.~\ref{fig:fig2}(a2), was characterized by the Dirac gap size of 58~meV. Figures~\ref{fig:5}(b1-b3) demonstrate the results of similar calculations for the case when the SOC strength was varied for all Bi and Te atoms within the two surface SLs. The resulting changes in the DP gap  size under modulation of the SOC coefficient for surface atoms in the range from 0 to 1.2 are presented in Fig.~\ref{fig:5}(d) by blue and red symbols for the SOC varied in the first SL and in the two SLs, respectively. A complete set of calculated dispersions in a wider energy range is shown in Fig.~4S and 5S of Suppl. Mater.

The contributions of the orbitals localized in the first and second/third SLs are shown in Fig.~\ref{fig:5}(a1-a4) and \ref{fig:5}(b1-b3) in red and blue, respectively (as before). The analysis shows that a decrease of the surface SOC strength coefficient from 1 to 0.8–0.9 leads to a sharp decrease in the DP gap width down to 5 meV and 28 meV for the systems with SOC modified in one and two SLs, respectively. With a further decrease in the SOC coefficient, a gradual increase in the DP gap size is observed up to 20~meV and 40~meV, respectively. For the values of $\lambda_{SOC}<1.1$ both upper and lower parts of the Dirac cone are clearly distinguishable among all band. At the same time, at $\lambda_{SOC}<0.9$ for both systems a noticeable deviation of the predicted splittings of the Te~$p_z$ states from the experimental values are observed. With an increase in the surface SOC coefficient up to 1.1 for both systems, our calculation predicts an increase of the Dirac gap width to 98~meV and 88~meV, respectively. Upon further increase of the SOC strength  up to 1.3–1.4 (see Figs.~4S and 5S in Suppl. Mater.) the states of the lower Dirac cone become already located below the edge of the VB states. As a result, the TSSs and VB states are difficult to separate. Thus, the presented estimates show that the DP gap may be varied in total from 5~meV to 98~meV with the SOC modulation for atoms in the first SL and from 28~meV to 88~meV with the SOC modulation for atoms in two surface SLs.

The obtained results can be compared with the results of similar calculations where the surface SOC was varied in the first and first two SLs, but for the initial structure with the bulk value of the surface SvdW (0~\%), which is characterized by the gap size of 29~meV at $\lambda_{SOC}=1$. These results are shown in Figs.~6S and 7S in Suppl. Mater. However, in these cases, the calculated splitting of the Te~$p_z$ states is greatly underestimated in comparison with the experiment, and the obtained values of the DP gap do not correlate well with the experiment.

For comparison, Fig.~\ref{fig:5}(c) shows the results of calculations for the system, in which a half of Bi atoms in the surface SL are replaced by Sb, which has lower atomic number. This corresponds to a decrease in the effective SOC for surface atoms. Calculations show that for such a system the size of the DP gap  decreases to 24~meV (compared to 58~meV), which is consistent with the results discussed above. 

\begin{figure*}
\includegraphics[scale=0.7]{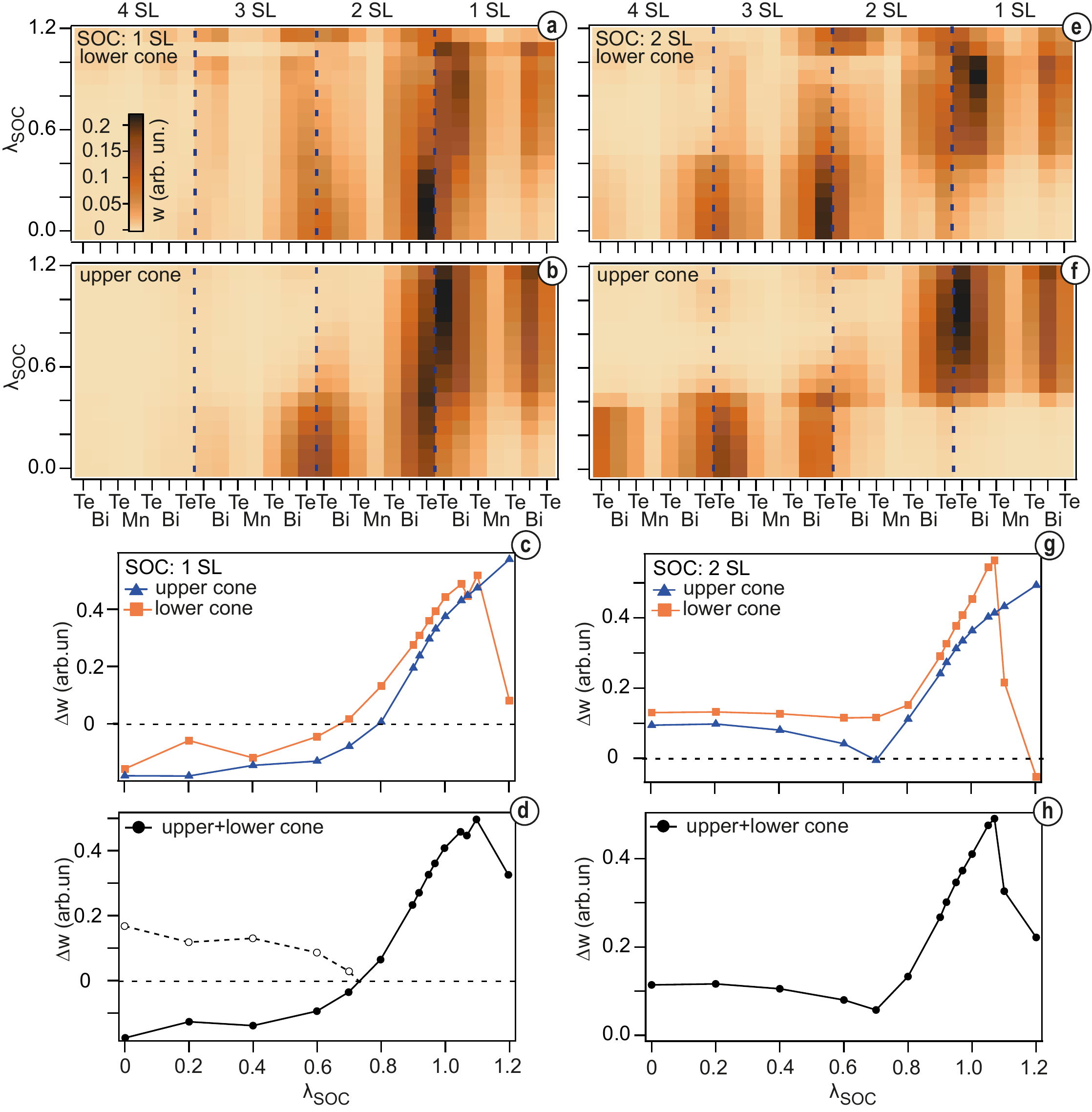}
\caption{\label{fig:6}(a,b and c,d) -- Redistribution in the TSSs density for the lower and upper parts of the Dirac cone in the region of four surface SLs with modulation of the SOC on Te and Bi atoms in one and two surface SLs (SOC:1SL and SOC:2SL systems). Dark color corresponds to the maximal TSSs density. (c,g) -- Corresponding changes in the signed TSSs density averaged over four surface SLs for the states of the lower and upper parts of the Dirac cone (blue and orange symbols) taking place under variation of SOC on Te and Bi atoms in one and two surface SLs. (d,h) -- Corresponding changes in the signed TSSs density averaged over the upper and lower Dirac cone states.}
\end{figure*}

To refine the picture of the changes in the size of the Dirac gap under modulation of the surface SOC, we calculated the corresponding changes in the localization of the TSSs. Figures~\ref{fig:6}(a,b) and \ref{fig:6}(e,f) show the changes in the spatial distribution of the lower and upper Dirac cone states in the regions of the first four SLs, depending on the variation in the SOC coefficient in one and two SLs, respectively. The calculations were carried out on the basis of the system with the Dirac gap of 58~meV. High local density in the TSSs distribution corresponds to the dark color, similarly to Fig.~\ref{fig:3}. Figures~\ref{fig:6}(a,b) and \ref{fig:6}(e,f) demonstrate that a decrease in the surface SOC coefficient below 1 leads to the shift of the TSSs localization towards lower-lying SLs, first to the area of the second/third SLs, and then, at values of the SOC coefficient less than 0.8, already to the area of the third and fourth SL for the system with SOC modified in one and two layers, respectively. On the other hand, with an increase of $\lambda_{SOC}$ from 1.0 to 1.1, the TSSs become localized mostly in the first/second SLs, this system is characterized by the maximum gap at the DP. When the SOC coefficient exceeds 1.1, the TSSs localization again shifts to the region of the second/third SLs. Thus, the maximum of the Dirac gap is reached when both the upper and the lower parts of the Dirac cone are mainly localized in the first and partially the second SLs. If we consider the DP gap widths for the two systems (SOC:1SL and SOC:2SL) in Fig.~\ref{fig:5}(d) and compare with the localization of TSSs in Figs.~\ref{fig:6}(a-f), we can see that the minimum of the gap width roughly corresponds to the situation when the sum of TSSs densities in the first and third SLs is equal to the sum of densities in the second and fourth SL. In this case the exchange fields produced by each SL cancel each other. Thus, the changes in the DP gap upon SOC variation are also based on the redistribution in the TSSs local density. 

As before, as a quantitative indicator of the effect of the TSSs redistribution on the Dirac gap size, we consider the TSSs density signed in accordance with the magnetization direction of each SL and averaged over all atomic layers. Figs.~\ref{fig:6}(c,g) show the changes in such average TSSs density that occur under the modulation of the SOC strength for atoms in one (Fig.~\ref{fig:6}(c)) and two (Fig.~\ref{fig:6}(g)) surface SLs. The data are presented separately for the upper and lower parts of the Dirac cone states (blue and orange symbols, respectively). Averaging was carried out over four SLs, taking into account the opposite signs of the magnetic moments in neighbor SLs. Figures~\ref{fig:6}(d,h) demonstrate the corresponding total changes in the average TSSs density summed over the upper and lower Dirac cone states. Taking into account the change in sign of the average TSSs density, the presented dependencies show a good correlation with the corresponding changes in the Dirac gap in Fig.~\ref{fig:5}(d). The gap minimum at $\lambda_{SOC}=0.8$ for the system SOC:1SL in ~\ref{fig:5}(d) corresponds to the sign change of the average TSSs density in Figs.~\ref{fig:6}(c,d). In this case, an increase in the DP gap size with decrease in the SOC strength below 0.8 also correlates with an increase in the absolute value of the average TSSs density (dashed line in Fig.~\ref{fig:6}(d)). In this case, the size of the gap is also determined by the the emerging average exchange field. For the system SOC:2SL Fig.~\ref{fig:6}(h) a good correlation with the gap width is also observed, although the TSSs density does not change its sign.

Thus, the presented results demonstrate that in the case of surface SOC modulation the changes in the TSSs localization are primarily responsible for the modulation of the DP gap. In this case, the non-monotonic character of the changes in the Dirac gap size is related either to non-monotonic variations or to the sign reversal of the average TSSs density, which is qualitatively related to the effective exchange field acting on the TSSs.

\begin{figure*}
\includegraphics[scale=0.6]{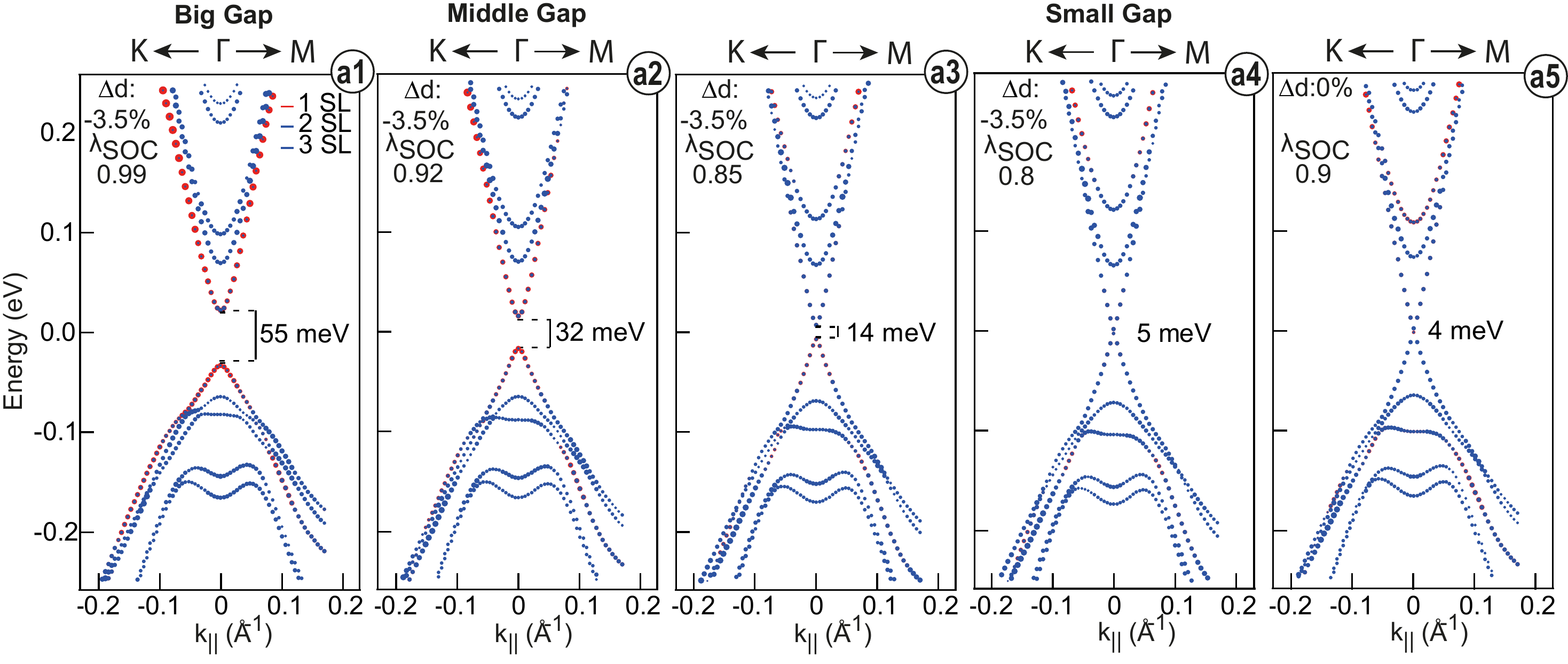}
\caption{\label{fig:7} Calculated dispersion maps for the TSSs and nearest VB and CB states, most corresponding to the experimental data, for a large (a1), medium (a2) and small (a3-a5) gap with the properly chosen calculation parameters (the SvdW interval between the first and second SLs and the SOC strength for the atoms in the first surface SL.}
\end{figure*}

\subsection{Comparative analysis of the calculated and experimental band dispersions}

To find conditions for the best agreement between the calculation results and experimental data, we show in Fig.~\ref{fig:7} several calculated dispersion maps at different parameters that provide good agreement with the experimental spectra presented in Fig.~\ref{fig:fig1}. A wider set of results is given in Fig.~8S. of Suppl. Mater. 

Samples with a large experimentally measured Dirac gap (55~meV) are best matched by the calculated systems characterized by a slightly compressed SvdW interval (-3.5~\% compared to the bulk) with the resulting gap size of 55-58~meV presented Fig.~\ref{fig:7}(a1) and Fig.~\ref{fig:fig2}(a2). In this case, the splitting of the Te~$p_z$ states correlates well with the experimental data. Wherein, some increase in the size of the gap to 77-80~meV is also possible due to additional decrease in the SvdW interval to –6~\% (see Fig.~\ref{fig:fig2}(a1)) as well as a slight modulation of the surface SOC in the first SL for the case of the SvdW interval of –3.46~\% (see Fig.~\ref{fig:5}(a4)). 

For samples with a middle experimentally measured gap (of about 28–32~meV \cite{shikin2021sample,shikin2022modulation}), the best agreement again takes place for the systems  with the compression of the subsurface SvdW interval by –3.46~\%, but also with a slightly modulated SOC for atoms in the first SL ($\lambda_{SOC}=0.9-0.92$), see Fig.~\ref{fig:7}(a2), as well as Fig.~8S. in Suppl. Mater. In this case, the splitting of the Te~$p_z$ states is also close to the experimental value.

For samples with a small measured gap (below 15-18~meV), the best agreement also takes place for systems with a  SvdW interval compressed by -3.46~\% and a more significant modulation of the surface SOC ($\lambda_{SOC}=0.85-0.9$), see Fig.~\ref{fig:7}(a3). The size of the Dirac gap in this case corresponds to 15~meV. With a further decrease in the surface SOC coefficient to 0.8, the gap size can reach the minimum values of 5~meV, see Fig.~\ref{fig:7}(a4). Interestingly, if we use the value of the SvdW interval corresponding to the bulk one (0~\%), then the gap can also reach values of 14~meV (at SOC=0.95) and 4~meV (at SOC=0.9), see Fig.~\ref{fig:7}(a5) for comparison, as well as Fig.~6S and 8S in Suppl. Mater. 

\section{Conclusions}

We analyzed possible changes in the electronic structure of the TSSs and the nearest VB and CB states for the AFM TI MnBi$_{2}$Te$_{4}$ with variations in the following calculation parameters: (i) the subsurface SvdW interval, (ii) the magnetic moment of Mn atoms, and (iii) the SOC strength for Te and Bi atoms in the first two SLs. It was found that all these factors can be responsible for the gap variation in the wide range between 55-60~meV (and even up to 80~meV) and 4-5~meV. In all these cases a significant redistribution of the TSSs takes place that mainly determines the Dirac gap modulation. In this case, the calculated non-monotonic character of the change in the Dirac gap size can be related either to similar behavior of the appropriately signed and averaged TSSs local density of states or to a change in its sign. This average TSSs density is qualitatively related to the effective exchange field acting on the TSSs.

It was found that the best agreement between the experimentally measured ARPES dispersion maps for all observed sizes of the Dirac gap (large, middle and small) and the results of calculations is reached with a slightly compressed SvdW interval between the first and second surface SLs (by -3.46~\% compared to the bulk). In the case of a middle and small gap the best agreement is reached under additional defect-induced modulation of the effective SOC for surface atoms in the first SL (about 0.9 for the middle gap and 0.8–0.85 for the small one). Although, a small gap can be also achieved in the calculations using the bulk value of the SvdW interval (0~\%) assuming the SOC modulation for surface atoms $\lambda_{SOC}=0.9$. In real systems, modulation of the Mn magnetic moments may be caused by Mn substitutions with nonmagnetic atoms, while changes in SOC can be related to substitution of Te and Bi atoms with other atoms providing different SOC. Our results demonstrate that such defect may lead to significant modifications of the DP gap width.

\section*{Data availability}

The authors declare that the data supporting the findings of this study are available within the paper.

\begin{acknowledgments}
The authors acknowledge support by Russian Science Foundation (Grant No. 18-12-00062) and the Saint Petersburg State University (Grant No. ID 90383050). The authors thank N. L. Zaytsev for helpful discussion.
\end{acknowledgments}

\providecommand{\noopsort}[1]{}\providecommand{\singleletter}[1]{#1}%
%


\end{document}